\documentclass[conference]{IEEEtran}
\IEEEoverridecommandlockouts
\usepackage{cite}
\usepackage{amsmath,amssymb,amsfonts}
\usepackage{algorithmic}
\usepackage{graphicx}
\usepackage{booktabs}
\usepackage{xcolor}
\usepackage{subfigure}
\usepackage{stfloats}
\usepackage{textcomp}
\def\BibTeX{{\rm B\kern-.05em{\sc i\kern-.025em b}\kern-.08em
    T\kern-.1667em\lower.7ex\hbox{E}\kern-.125emX}}
\begin{document}

\title{A Deep Deterministic Policy Gradient-based Strategy for Stocks Portfolio Management}

\author{\IEEEauthorblockN{Huanming Zhang}
\IEEEauthorblockA{\textit{Dept. of Mathematical Sciences,} \\
\textit{University of Liverpool}\\
Liverpool, United Kingdom }
\and
\IEEEauthorblockN{Zhengyong Jiang}
\IEEEauthorblockA{\textit{Dept. of Mathematical Sciences,} \\
\textit{Xi’an Jiaotong-Liverpool University}\\
Suzhou, P.R.China \\
Zhengyong.Jiang@xjtlu.edu.cn}
\and
\IEEEauthorblockN{Jionglong Su}
\IEEEauthorblockA{\textit{School of AI and Advanced Computing,} \\
\textit{XJTLU Entrepreneur College (Taicang),}\\
\textit{Xi’an Jiaotong-Liverpool University}\\
Suzhou, P.R.China \\
Jionglong.Su@xjtlu.edu.cn}
}

\maketitle

\begin{abstract}
With the improvement of computer performance and the development of GPU-accelerated technology, trading with machine learning algorithms has attracted the attention of many researchers and practitioners. In this research, we propose a novel portfolio management strategy based on the framework of Deep Deterministic Policy Gradient, a policy-based reinforcement learning framework, and compare its performance to that of other trading strategies. In our framework, two Long Short-Term Memory neural networks and two fully connected neural networks are constructed. We also investigate the performance of our strategy with and without transaction costs. Experimentally, we choose eight US stocks consisting of four low-volatility stocks and four high-volatility stocks. We compare the compound annual return rate of our strategy against seven other strategies, e.g., Uniform Buy and Hold, Exponential Gradient and Universal Portfolios. In our case, the compound annual return rate is 14.12\%, outperforming all other strategies. Furthermore, in terms of Sharpe Ratio (0.5988), our strategy is nearly 33\% higher than that of the second-best performing strategy. 
\end{abstract}


\section{Introduction}
Portfolio management is generally defined as the process of allocating assets such as stocks, futures and options  by constructing an internal representation (model) of the markets\cite{b1}.
Based on existing theories, many timing strategies, stock-picking strategies, statistic arbitrage and portfolio management strategies are proposed in the last decades. For example, Dzitkowski and Kozicki employed Long Short-Term Memory (LSTM) in High Frequency Trading\cite{b2}; Mehtab, Sen and Dutta analyzed the sentiments in social media and utilized the sentiment-related information based on self-organizing fuzzy neural networks (SOFNN) to predict stock price\cite{b3}. Such supervised-learning algorithms rely on historical information, e.g., price information, calculated indices, non-financial information and other information to predict future price movements. Unfortunately, given the efﬁcient market hypothesis and non-stationarity property in the price time series, it is seemingly impossible to truly make price predictions timely and accurately. Furthermore, price prediction does not translate directly into actual market action.  In addition to the prediction of price movement, rational judgment is also required to implement portfolio strategy. 

Recently, many studies on the use of reinforcement learning have drawn much attention of research not only because they no longer need to predict price, but due to their superior performance in Board Games, Go, and Video Games\cite{b4}, which mimic the environment of financial market to some extent. Gao et al.\cite{b5} \cite{b24} discretized the action space which is defined as the weight of portfolio wealth in different stocks, and combined the Convolutional Neural Network (CNN) with dueling Q-net (DQN) to obtain the optimal actions. However, finite action space models might result in local optimum or no optimum at all, as they are limited by the number of action combinations. Besides, Gao et al.'s strategy model\cite{b5} \cite{b24} and many other traditional strategies such as Exponential Gradient (EG)\cite{b20} and Online Moving Average Reversion (OLMAR)\cite{b21} dismiss transaction cost. Apart from the above limitations, some strategies are simply tested in a short period of time, e.g., spanning one or two years, which makes it difficult to experience different market phases.

In this paper, instead of discretizing the continuous action space into finite action space, we use a Deep Deterministic Policy Gradient (DDPG)\cite{b4} framework specifically designed for multi-asset trading strategy in continuous action space. This framework allows the DDPG agent to receive the market states and then optimize trading policies (actions), which are the portfolio wealth allocations in different time steps. The framework will output the optimized trading policies (actions) for each time step, which we will subsequently use to calculate the returns and other performance metrics to evaluate our DDPG-based strategy.

The motivation of our work is two-fold. First, the need of applying reinforcement learning model in continuous action space in portfolio management inspires us to apply the DDPG framework. Second, different from previous work which dismisses transaction cost\cite{b5} \cite{b17} \cite{b21} \cite{b19}, we want to explore the impact of transaction cost on our strategy. 

The novelty of our work is as follows: Unlike other DDPG-based models where CNN are employed as both critics and actors network\cite{b5}\cite{b10}, we propose a new framework, which consists of two LSTM networks working as critic and target-critic network respectively, as well as two fully connected (FC) networks working as actor and target-actor.

The key contribution of our research is that we compare our DDPG-based strategy's performance under the conditions with and without transaction cost, and discover a counterintuitive result that the strategy with transaction cost is more robust and profitable than that without transaction cost, we also provide a plausible explanation for this phenomenon in our experiment. Furthermore, we assess the efficiency of our strategy against seven other traditional strategies using the same dataset, and find their returns are neither as stable nor as high as our DDPG strategy. The compound annual return rate of our DDPG strategy (14.12\%) is at least twice as much as other traditional strategies. In terms of Sharpe Ratio (SR), our DDPG strategy (0.5988) is nearly 33\% higher than that of the second best strategy (0.3679). 

The organization of the rest of this paper is as follows. Section \uppercase\expandafter{\romannumeral2} describes the dataset we use. Section \uppercase\expandafter{\romannumeral3} provides a mathematical treatment to our problem, and also gives a brief introduction to our DDPG strategy in comparison to other traditional RL methods.
Section \uppercase\expandafter{\romannumeral4} proposes some basic assumptions used in our research, Section \uppercase\expandafter{\romannumeral5} presents a detailed description of our strategy. Section \uppercase\expandafter{\romannumeral6} contains the results as well as the analysis of our strategy. Finally in Section \uppercase\expandafter{\romannumeral7},  conclusions of the research and future research directions are given.

\section{Dataset and Features}

The stock data comes from Yahoo Finance including eight US stocks ranging from July 1999 to July 2020. Notably, the data includes the 2008 market crash as well as 2020 Covid-19 market crisis in order to train and test our model with real world fluctuations. We choose four stocks with low volatility measured by $\mathit{Beta}$ $\mathit{Index}$ and four stocks with high volatility to formulate our portfolio. Four high volatility stocks are: $\mathbf{APA}$, $\mathbf{LNC}$, $\mathbf{RCL}$, $\mathbf{FCX}$\cite{b7}(see Figure 1(a)); four low volatility stocks are: $\mathbf{GOLD}$, $\mathbf{FDP}$, $\mathbf{NEM}$, $\mathbf{BMY}$\cite{b8}(see Figure 1(b)). 

For data pre-processing, we calculate the Relative Strength Index of two-day time period (RSI2), daily simple return rate and logarithm daily return rate for each stock. Furthermore, we directly drop off the empty values since the price information for some stocks are not available every day, but such situation is rare, which means directly omitting empty values will not significantly influence our result.

\begin{figure*}[tbp]
	
	\subfigure{\includegraphics[width=0.5\textwidth]{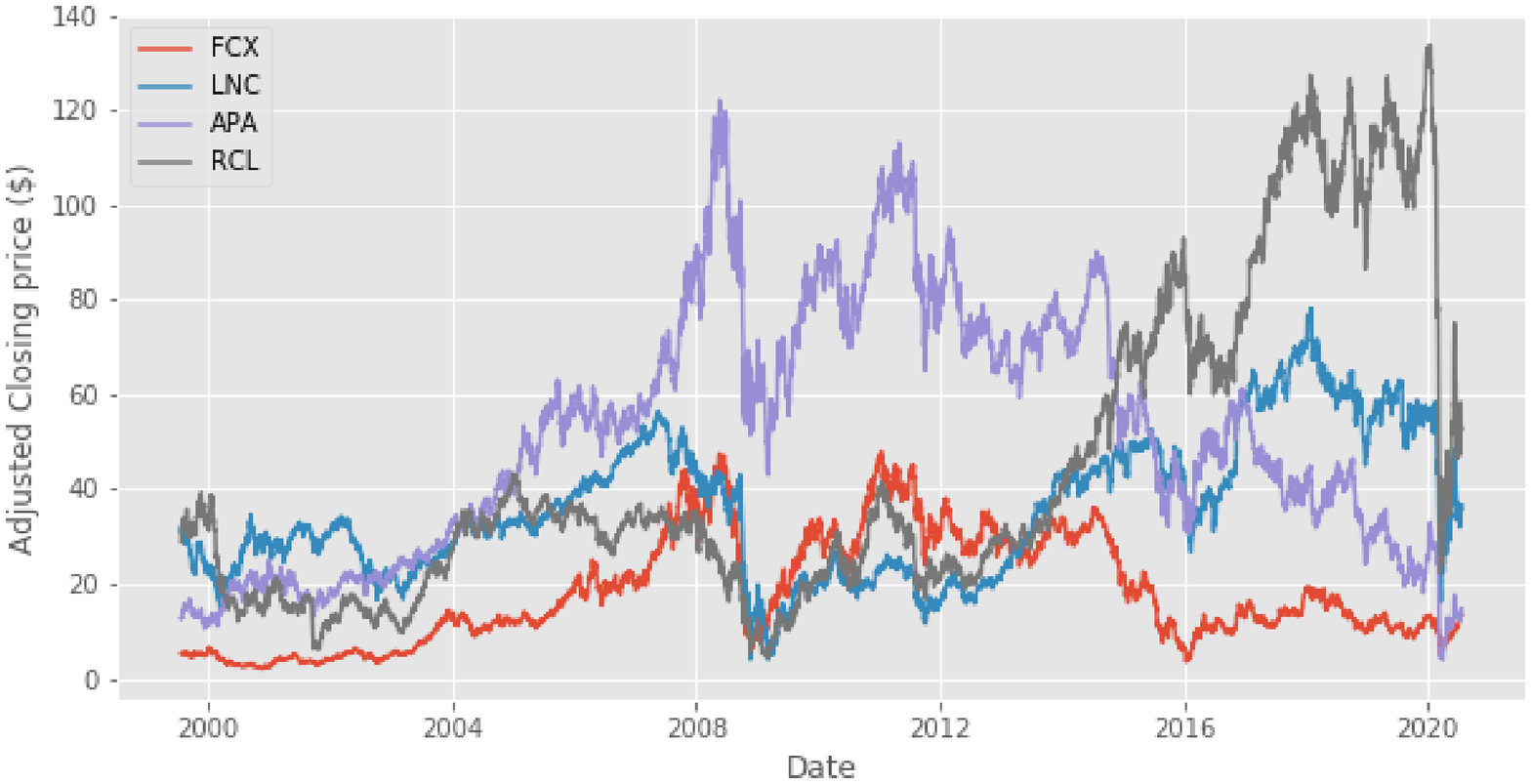}}
	\subfigure{\includegraphics[width=0.5\textwidth]{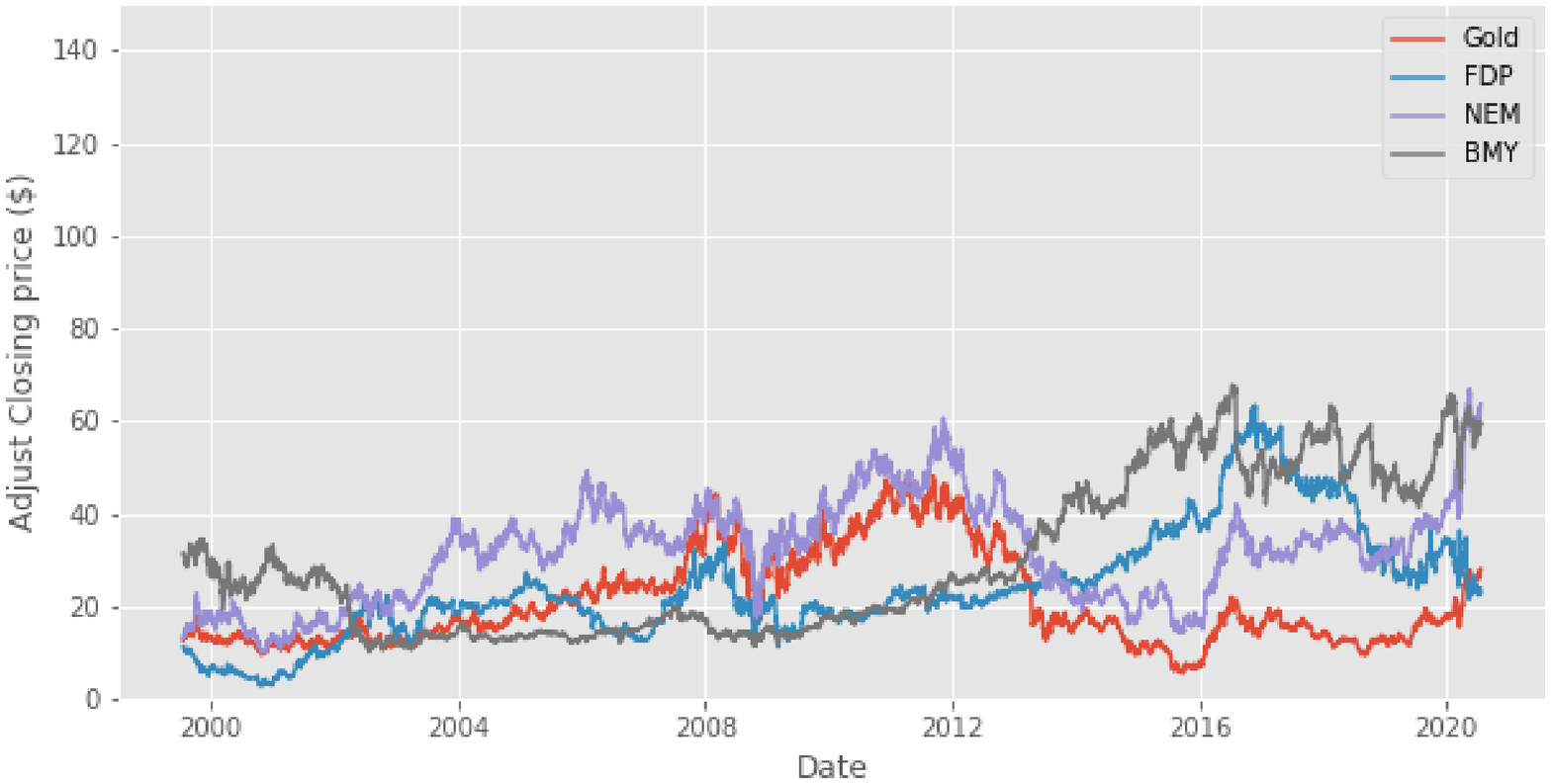}}
	\caption{Daily Stocks Closing Price From July 1999 to July 2020  \protect\\
		(a)Adjusted Prices of High Volatility Stocks;  \hspace{4cm}
		(b)Adjusted Prices of Low Volatility Stocks}
	\label{fig:1}
\end{figure*}

\section{Problem Statement}

Portfolio management is the action of continuous reallocation of a capital into a number of ﬁnancial assets\cite{b23}. In practice, the investors modify the weight of portfolio in different assets at each trade period. In this work, we set the trade period as one day. We shall first provide some mathematical notations:

\begin{itemize}
	\item $M$: The number of assets, here we have $M = 8$
	\item $\boldsymbol{w}_t$\cite{b1}: Portfolio vector, its $i$-th component represents the ratio of the total budget invested to the $i$-th asset at time $t$, such that
	\[
	\begin{split}
	\boldsymbol{w}_{t}=\left[w_{1, t}, w_{2, t}, \ldots, w_{M, t}\right]^{T} \in \mathbb{R}^{M} 
	\text { and }  \sum_{i=1}^{M} w_{i, t}=1 \\
	w_{i,t} \in \left[0,1\right]
	\end{split}
	\]
	\item $p_{i,t}$: The adjusted closing price of stock $i$ at time $t$ 
	\item $\text{RSI2}_{i,t}$: The RSI of two-day time period of stock $i$ at time $t$
	\item $h_{i,t}$: The shares holding of stock $i$ at time $t$
	$$
	h_{i,t} = \left \lfloor \frac{V_t \times w_{i, t}}{p_{i,t}} \right \rfloor
	$$
	where $\lfloor x \rfloor$ means the largest integer less or equal than $x$
	\item $V_t$: Portfolio Value at time $t$
	$$
	V_t = \sum_{i = 1}^M h_{i,t-1}\times p_{i,t} + c_{t-1}
	$$
	\item $c_t$: Cash owned at time $t$
	
	\item $\boldsymbol{s}_t$: State at time $t$, a tuple consisting of $\left(p_{i,t}, p_{i, t-1}, \log{\frac{p_{i,t}}{p_{i, t- 1}}}, \text{RSI2}_{i,t}, h_{i,t}, V_t, c_t\right)$, where $i = 1,2...,8$ and $t \geqslant 1$, thus $\boldsymbol{s}_t \in \mathbb{R}^{42}$
	\item $R_t$: Reward at time $t$, defined as:
	$$
	R_{t} \triangleq V_{t}-V_{t-1}
	$$ 
	\item $\boldsymbol{a}_t$: Action taken at time $t$, we treat it as the the ratio of total budget of each asset, i.e.
	$$
	\boldsymbol{a}_{t} \triangleq \boldsymbol{w}_{t+1}
	$$
	\item $\pi_t(a | s)$:  The probability that $\boldsymbol{a}_t = \boldsymbol{a}$ if $\boldsymbol{s}_t = \boldsymbol{s}$. 
	\item $q_{\pi}(s,a)$: The value of taking action $\boldsymbol{a}$ in state $s$ under a policy $\pi$
	
	\[
	\begin{split}
	q_{\pi}(s, a)&=\mathbb{E}_{\pi}\left[G_{t} \mid S_{t}=s, A_{t}=a\right] \\
	&=\mathbb{E}_{\pi}\left[\sum_{k=0}^{\infty} \gamma^{k} R_{t+k+1} \mid S_{t}=s, A_{t}=a\right]
	\end{split}
	\]
	where $\gamma \in \left(0,1\right]$ is called discounting factor 
\end{itemize}
Traditionally, the reinforcement learning problem is a straightforward framing of the problem of learning from interaction between agent and environment to achieve a goal. The learner and decision-maker is called the agent, while the environment is the outside thing it interacts with. These interact continuously, i.e., the agent chooses actions, and the environment responds to the actions and return rewards that the agent seeks to maximize overtime. To achieve this, the agent needs to modify its action following policy, a mapping from states to probabilities of selecting each possible action\cite{b9}.

If both the action space and state space are not large, rendering a possibility to record the $q$ value of all combination of states $s$ and actions $a$, the agent can determine its optimal action by looking up the $q$ value from $q$-table, i.e. 
$$
\pi_{*}(a \mid s)=\left\{\begin{array}{ll}
1 & \text { if } a=\underset{a \in A}{\operatorname{argmax}}\quad q_{*}(s, a) \\
0 & \text { otherwise }
\end{array}\right.
$$  

\begin{figure}[htbp]
	\centerline{\includegraphics[width=0.5\textwidth]{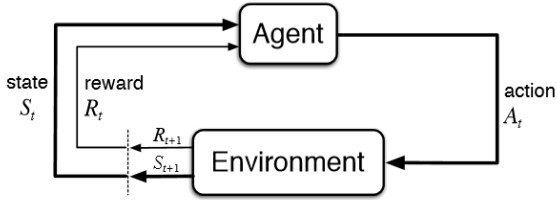}}
	\caption{The agent–environment interaction in reinforcement learning\cite{b9}}
	\label{fig:2}
\end{figure}

In our experiment, however, the state space is of 42-dimension, and each entry of one state is a continuous real number, and so as the action space. Therefore, it is neither feasible nor reasonable to use table-lookup method to obtain the optimal policy. We shall use DDPG as our basic model because it does not have special requirements for state space and action space, which means it can solve either finite or infinite space problems.

\section{Assumptions}
\begin{itemize}
	\item Assumption 1:
	The liquidity of all market assets is high enough that, each trade can be executed immediately at the orderd price, i.e. zero slippage. 
	\item Assumption 2: The action of agent will not affect the market.
	\item Assumption 3: A small transaction cost per transaction (\$0.001) was used to encapsulate the various trading fees\cite{b6}, and no other commisions.
	\item Assumption 4: The portfolio keeps unchanged during the period from the end of previous trading day and the beginning of next trading day.
	\item Assumption 5: No other assets are chosen besides the selected stocks.
	\item Assumption 6: The volume of each stock is large enough, so the model can buy or sell every stock at any trading day.
\end{itemize}

\section{Methods}
\subsection{From Actor-Critics Methods to DDPG}

Actor–critic methods\cite{b4} are temporal-difference (TD) methods with a separate memory structure to explicitly represent the policy without depending on the value function. The policy structure is known as the actor, since it is used to select actions given the input state. The estimated value function is known as the critic as it evaluates the action value made by the actor. The critic learns about and critique the policy being adopted by the actor. The critique takes the form of a TD error\cite{b9}:  
$$
\delta_{t}=R_{t+1}+\gamma V\left(S_{t+1}\right)-V\left(S_{t}\right)
$$
where $V$ is the current value function implemented by the critic.

\begin{figure}[htbp]
	\centerline{\includegraphics[width=0.3\textwidth]{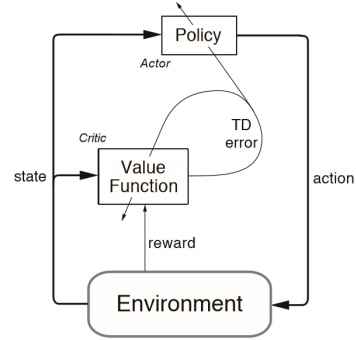}}
	\caption{The actor–critic architecture\cite{b9}}
	\label{fig:3}
\end{figure}

Typically, the critic is a function with state as input and an evaluation value as output. After each action is selected, the critic evaluates the new state to see if things turn out to be better or worse than expected. Similarly, following the feedback from critic, the actor learns how valuable its action is, and tends to choose the optimal action.

Essentially, both the critic and actor are function approximators, and therefore we can use neural network to construct them. The DDPG algorithm (Figure 4) can generate deterministic actions directly from the policy of actor without sampling according to the probability distribution of action, so it is called deterministic policy. In the learning phase, a noise function is added to the deterministic action to realize the small-scale exploration near such deterministic action\cite{b11}. In addition, the algorithm also backups a set of parameters for actor and critical network respectively to calculate the expectation of action value, so as to improve the guidance level of critics\cite{b4}. The networks using the backup parameters are called the target networks. The purpose of such two sets parameter design (backup parameters and real-time updating parameters) is to reduce the non-convergence caused by the approximated data\cite{b9}.

\begin{figure}[htbp]
	\centerline{\includegraphics[width=0.5\textwidth]{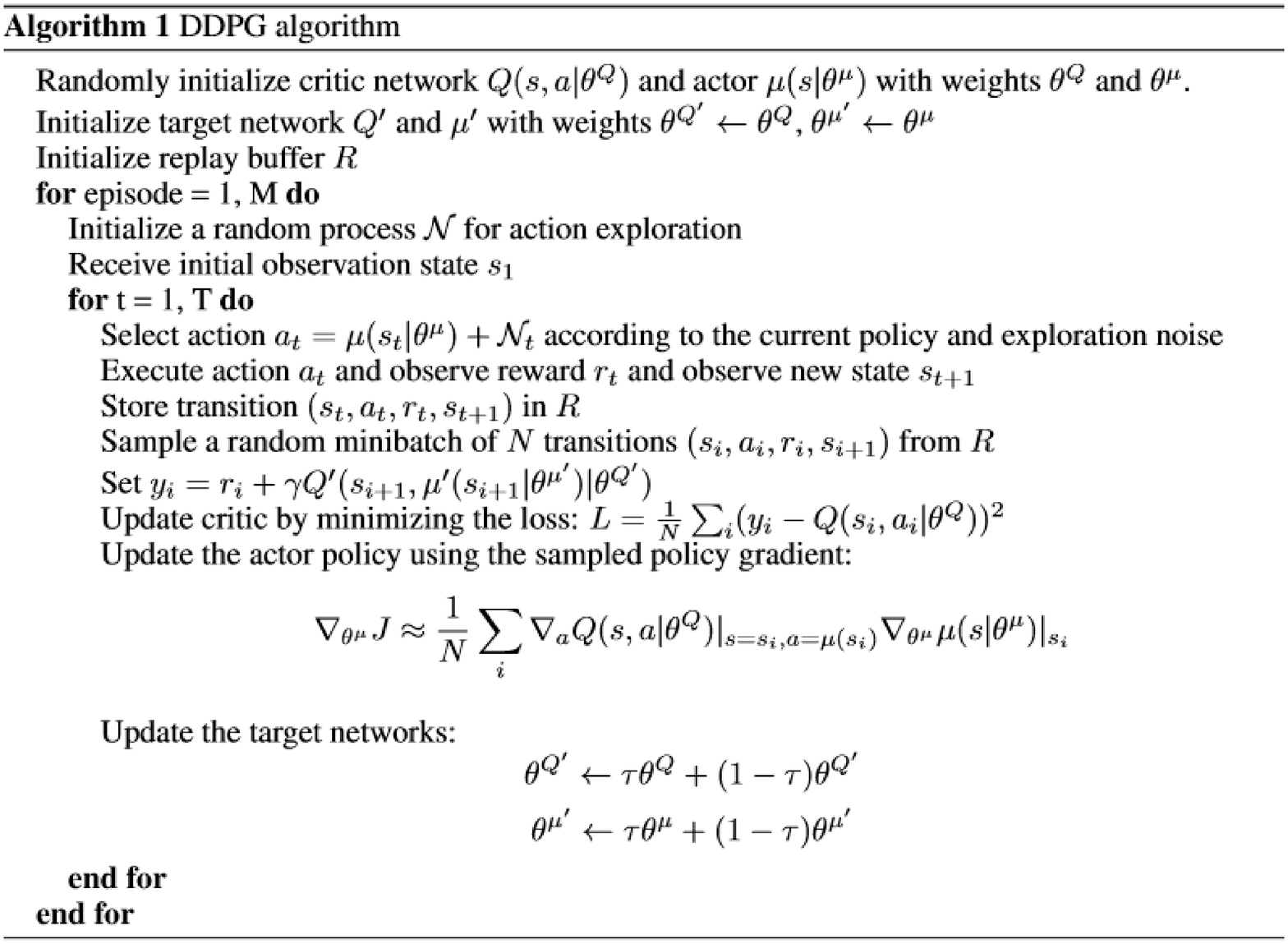}}
	\caption{DDPG Algorithm\cite{b11}}
	\label{fig:4}
\end{figure}

The specific application scenarios of these four networks are as follows:
\begin{itemize}
	\item Actor Network: According to the current state $s_0$, generating specific action $a_0$ of exploring or not exploring.
	\item Target-Actor Network: Based on the follow-up state $s_1$ given by the environment, a new action $a_1$ is generated.
	\item Critic Network: Computing state-action value $Q_{\text{predicted}}(s_0,a_0)$ 
	\item Target-Critic Network: Given subsequent state-action pair $(s_1, a_1)$, calculating $Q(s_1, a_1)$ to compute expected $Q$, i.e., $Q_{\text{expected}}(s_0, a_0) = R_1 + \gamma * Q(s_1, a_1)$
\end{itemize}

\subsection{Data Processing}
Since the inputs of critic network and target-critic network are state $\boldsymbol{s}_t \in \mathbb{R}^{42}$ and action $\boldsymbol{a}_t \in \mathbb{R}^{8}$, we concatenate these two vectors to form a new vector $\left[s_t, a_t\right] \in \mathbb{R}^{50}$.

\subsection{Network Frameworks}
For the critic network and target-critic network, we choose an LSTM network (figure 5) with 3 hidden layers for implementation. The LSTM is a promising choice to handle with time-series as it is effective in solving the problems regarding gradient vanish and gradient explosion.

\begin{figure}[htbp]
	\centerline{\includegraphics[width=0.4\textwidth]{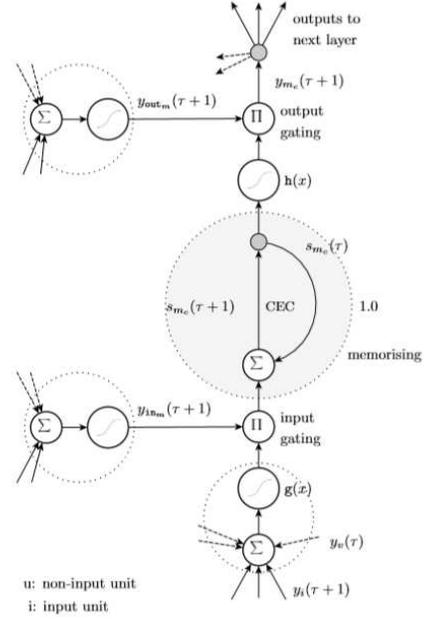}}
	\caption{A standard LSTM memory block\cite{b12}}
	\label{fig:5}
\end{figure}

The input dimension is $\left(\text{batch size}, \text{input size}\right) = \left(128, 50\right) $ since our mini-batch size is $128$. The hidden size (node number) of the first layer and second layer are $100$ respectively. We also set hyperparameter ``dropout" to 0.35. Following the first two LSTM layers, we construct a fully connnected layer with hidden size $50$, and finally an output layer showing $Q$ value(Figure 6). 

\begin{figure}[htbp]
	\centerline{\includegraphics[width=0.3\textwidth]{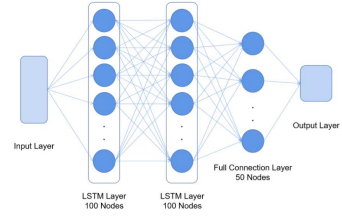}}
	\caption{Critic Network}
	\label{fig:6}
\end{figure}

To evalute the critic network, we use Huber-Loss with $\delta = 1$: 
$$
L_{\delta}(y, f(x))=\left\{\begin{array}{ll}
\frac{1}{2}(y-f(x))^{2} & \text { for }|y-f(x)| \leq \delta \\
\delta|y-f(x)|-\frac{1}{2} \delta^{2} & \text { otherwise }
\end{array}\right.
$$

\begin{figure}[htbp]
	\centerline{\includegraphics[width=0.4\textwidth]{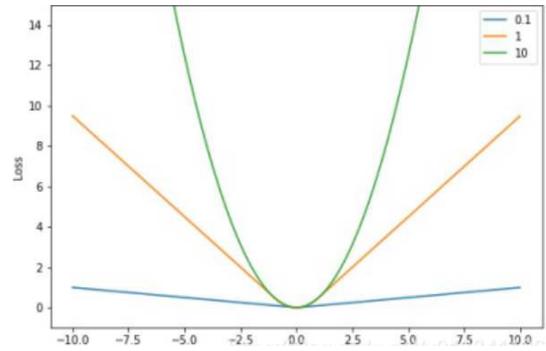}}
	\caption{Huber Loss with Different $\delta$}
	\label{fig:7}
\end{figure}

For actor network (Figure 8), we use fully connected (FC) network layout. The input is the state variable $\boldsymbol{s}_t$ of 42-dimensioned. The size of three hidden layers are $256, 128, 64$ respectively, with $\mathit{relu}$ function as activation. The output is an 8-dimensioned action vector $\boldsymbol{a}_t$ with each entry $a_t^{(i)} \in \left[0,1\right]$. Thus we use $\mathit{sigmoid}$ function to project the raw output to interval $\left[0,1\right]$. The loss function of actor network\cite{b4} is as follows:

$$
\text{loss} = -\sum_i^B Q_{\text{predicted}}\left(s_{t_i}, A\left(s_{t_i}\right)\right)
$$
where $A(s_t)$ is the output of actor network, $B$ is the mini-batch size, $Q_{\text{predicted}}$ is the output of critic network.

\begin{figure}[htbp]
	\centerline{\includegraphics[width=0.3\textwidth]{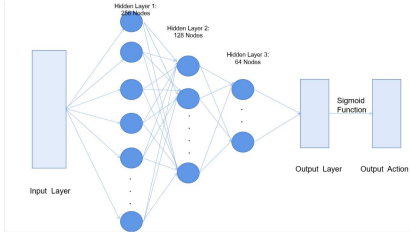}}
	\caption{Actor Network}
	\label{fig:8}
\end{figure}

\subsection{Tradeoff between Exploitation and Exploration}
We use $\epsilon$-greedy algorithm to provide a balance between exploitation and exploration. For each learning, the program generates a random number between 0 and 1, saying $r$, if $r > \epsilon$, we exploit, and if $r \leqslant\label{key} \epsilon$, we explore using Ornstein Uhlenbeck Action Noise (OU noise) algorithm\cite{b13} instead of Gaussian Noise, as OU noise is autocorrelated while Gaussian Noise are time-independent (Figure 9). In particular, we set a batch of stratified $\epsilon$ in our experiment: \\
We firstly denote as $N_t$ the total state number our model has experienced at time $t$.
$$
\epsilon_t=\left\{\begin{array}{ll}
0.5 & , \quad N_{t}<1000 \\
0.25, & 1000 \leq N_{t}<2000 \\
0.1, & 2000 \leq N_{t}
\end{array}\right.
$$
We also set the soft update hyperparameter $\tau$ to be 0.09.

\begin{figure}[htbp]
	\centerline{\includegraphics[width=0.4\textwidth]{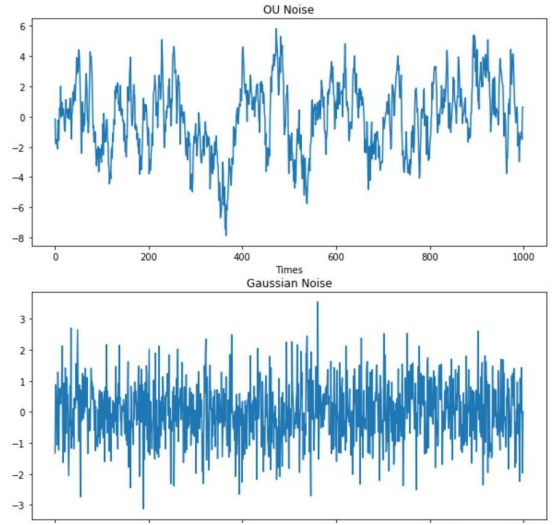}}
	\caption{OU Noise \& Gaussian Noise}
	\label{fig:9}
\end{figure}

\subsection{Performance Metrics}
Three different metrics have been used to evaluate the performance of trading strategies. The first is Compound Annual Return Rate (CARR), which is defined as: 
$$
C A R R=\left(\frac{E V}{B V}\right)^{\frac{1}{n}}-1
$$ 

where:
\begin{itemize}
	\item $\mathit{EV}$: The portfolio wealth at the last test day. 
	\item $\mathit{BV}$: The portfolio wealth at the begining test day.
	\item $\mathit{n}$: Number of years
\end{itemize}

The second metric is Sharpe Ratio (SR) which mainly represents the risk adjusted return of strategies. \cite{b15}: 
$$
S R=\frac{\mathbb{E}_{t}\left[\rho_{t}-\rho_{R F}\right]}{\sqrt{\operatorname{Var_t}\left(\rho_{t}-\rho_{R F}\right)}}
$$
where
$$
\rho_{t} \triangleq \frac{V_{t}}{V_{t-1}}-1
$$
and $\rho_{RF}$ represents the rate of return of risk-free asset. We set $\rho_{RF} = 0$ in our experiment.

The third metric is the Maximum Drawdown (MDD), which we use to assess the risk resistance of an investment strategy completely\cite{b16}. 
$$
MDD = \max_{t_2 > t_1} \frac{V_{t_1} - V_{t_2}}{V_{t_2}}
$$
where $t_1$ and $t_2$ represent different time steps. 

This metric denotes the maximum portfolio value loss from a peak to bottom. 

\section{Result and Analysis}

\subsection{Settings}
We set initial money to be 1,000,000, and the trading frequency is daily. In other words, we reallocate our portfolio everyday. We also choose the data from July 1999 to July 2016 as training set, and backtest our strategy in the whole dataset from July 1999 to July 2020. 

For a more intuitive comparison, we choose a benchmark strategy: At the first test day, we allocate the same amount of money to each stock, i.e. $\boldsymbol{w}_0 = \frac{1}{8} \left[1,1,1,1,1,1,1,1\right]$, then we hold these stocks shares until the end test day.

\subsection{Cost-Free DDPG Strategy}
We first give the result of the Cost-free DDPG strategy (Figure 10), in which we omit the transaction cost. From this figure, we observe that the performance of the Cost-free DDPG strategy from 1999 to the end of 2008 does not exceed that of the benchmark strategy. The performance of the two strategies is almost identical in this time period. However, starting from 2009, the Cost-free DDPG strategy performs significantly better than the benchmark strategy, even though it experiences larger drawdowns in 2015, 2018 and 2020. To some extent, the Cost-free DDPG strategy is an amplification of the benchmark strategy, both on return and loss. The Cost-free strategy's Compound Annual Return Rate (CARR) is 11.78\%, Sharpe Ratio (SR) is 0.4781, Maximum Drawdown (MDD) is 56.67\%.
\begin{figure}[htbp]
	\centerline{\includegraphics[width=0.53\textwidth]{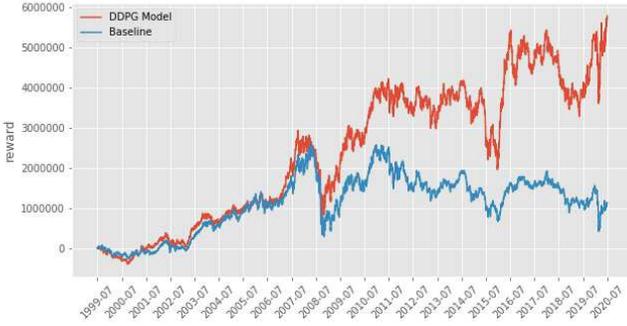}}
	\caption{Cost-free DDPG Strategy}
	\label{fig:10}
\end{figure}

\subsection{Cost DDPG Strategy}
The return curve of the Cost DDPG strategy is given in Figure 11, where we take into account of the transaction cost of \$0.001 per share for each transaction. Also, the profit distribution histogram is presented (Figure 12).

\begin{figure}[htbp]
	\centerline{\includegraphics[width=0.5\textwidth]{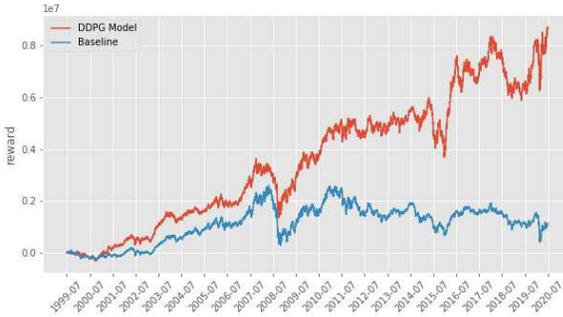}}
	\caption{Cost DDPG Strategy}
	\label{fig:11}
\end{figure}

Compared to the benchmark strategy, the return of Cost DDPG strategy outperforms that of benchmark from 2001, which is earlier than the Cost-free DDPG strategy. After 2001, the return of Cost DDPG strategy start to steadily increase until 2008, with the arrival of the global financial crisis. Both benchmark strategy and Cost DDPG strategy experience a pronounced drawdown, while the drawdown of Cost DDPG strategy is less than that of the benchmark. Since 2008, Cost DDPG strategy's return begins to rise again. If we particularly focus on the time period after July 2016, the return of the Cost DDPG strategy continuously increases until the beginning of 2020, as the Covid-19 epidemic deals a huge blow to the global market. Finally, the Cost DDPG strategy's Compound Annual Return Rate (CARR) is 14.12\%, Sharpe Ratio (SR) is 0.5988, Maximum Drawdown (MDD) is 49.13\%.

\begin{figure}[htbp]
	\centerline{\includegraphics[width=0.5\textwidth]{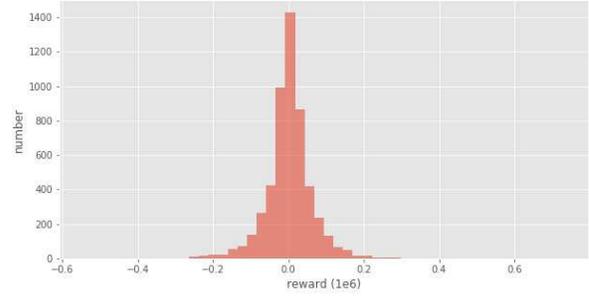}}
	\caption{Cost DDPG Strategy Profit Distribution}
	\label{fig:12}
\end{figure}

\subsection{Comparison Between Cost Strategy and Cost-free Strategy}
Both Cost-Free strategy and Cost strategy performed similarly in the first two years, so does the benchmark strategy. It is possible to explain this as both strategies underfitted at the initial stage since the training data is not sufficient enough. Starting from 2001, all three strategies increase steadily until 2008, when all strategies experience a significant drawdown, but the Cost DDPG strategy is the most robust during this period, with its drawdown being the smallest among the three. Subsequently, the returns of all three strategies rebound to varying degrees, but the returns of both the Cost-free DDPG strategy and the benchmark strategy remain volatile after 2011. When we specifically focus on time period after 2016, both Cost-free and Cost DDPG strategies make profits during this period, despite their returns declining sharply in 2020 due to Covid-19. 

One counterintuitive result is that the Cost DDPG strategy performs better than the Cost-Free DDPG strategy in our experiment. We should provide a plausible explanation here: The transaction cost added in reward function could be treated as a small constant penalty term, which will encourage our model to choose cautious actions, i.e. the actions making profits not only without transaction cost, but also with transaction cost. Moreover, a small amount of transaction costs can also make the strategy profitable without expending too much on transaction fees.

Jin and El-Saawy\cite{b6}'s work indicates that a penalty term added in reward might make the strategy more robust and profitable. This partly validates our explanation on why Cost DDPG strategy outperforms the Cost-free DDPG strategy. They modify the portfolio reward to include a penalty for volatility: 
$$
P_t = V_t - \lambda \sqrt{\mathrm{Var}(V_t)}
$$ 
where $V_t$ is the portfolio value at time $t$, $\lambda \in\left(0,1\right]$ is the penalty factor, and $P_t$ is the modified portfolio value at time $t$, $\mathrm{Var}$ is the variance. They test their Deep Q-Learning strategy (DQN) in different portfolio, and found out the penalized strategy are more stable even in different markets. Figure 13 shows their DQN strategy performance while choosing CPK and CHK as trading stocks, and Figure 14 presents the result in AVA and ETFC.

\begin{figure}[htbp]
	\centerline{\includegraphics[width=0.5\textwidth]{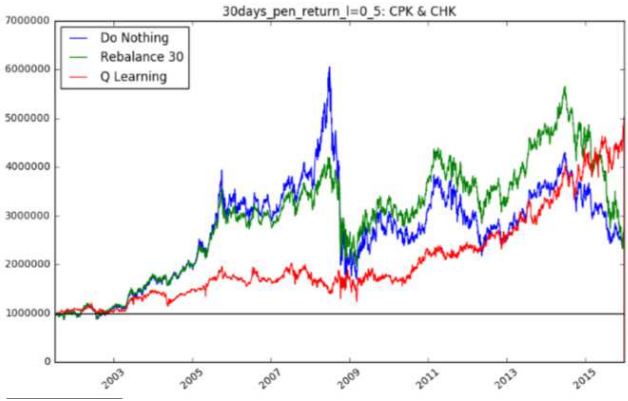}}
	\caption{DQN with Penalty Term in CPK \& CHK\cite{b6}}
	\label{fig:13}
\end{figure}

\begin{figure}[htbp]
	\centerline{\includegraphics[width=0.5\textwidth]{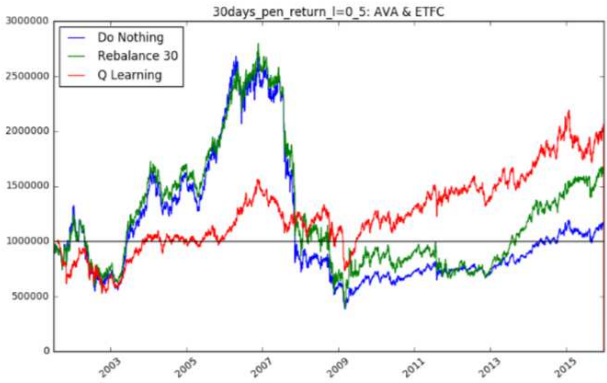}}
	\caption{DQN with Penalty Term in AVA \& ETFC\cite{b6}}
	\label{fig:14}
\end{figure}

\subsection{Strategy Comparison}
We compare our cost DDPG-based strategy with several traditional strategies as below:
\begin{itemize}
	\item Aniticor (ANTICOR) \cite{b17}
	\item The Uniform Buy and Hold (BAH), a portfolio management approach simply equally spreading the total wealth into the selected assets and holding them  without making any purchases or selling until the end. This is our benchmark strategy in previous graphs.\cite{b18}
	\item Uniform Constant Rebalanced Portfolios (CRP) \cite{b19}
	\item Exponential Gradient (EG) \cite{b20}
	\item Online Moving Average Reversion (OLMAR) \cite{b21}
	\item Passive Aggressive Mean Reversion (PAMR) \cite{b22}
	\item Universal Portfolios (UP) \cite{b19}
\end{itemize}

All the above strategies are tested with the same transaction cost as our cost DDPG-based strategy. Figure 15 illustrates the accumulative return over the investment horizon of the test period.

\begin{figure}[htbp]
	\centerline{\includegraphics[width=0.53\textwidth]{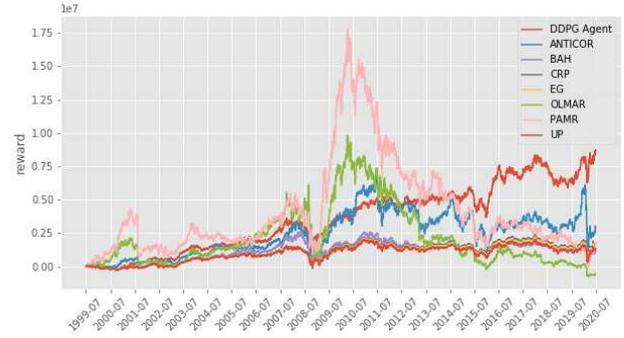}}
	\caption{Strategy Comparison}
	\label{fig:19}
\end{figure}

\begin{table}[!htbp]
	\centering
	\caption{Performance Metrics Comparison. The figures in bold are the best results for each metric}\label{tab:tab1}
	\begin{tabular}{cccc}
		\toprule
		& CARR& SR& MDD\\
		\midrule
		Anticor& 7.94\%& 0.3679 &0.8117\\
		BAH& 4.37\%& 0.2711& 0.6423\\
		CRP& 5.77\%& 0.3141& 0.6428\\
		EG& 5.39\%& 0.3027& 0.6415\\
		OLMAR& -4.06\%& 0.2331& 0.9780\\
		PAMR& 6.09\%& 0.3595& 0.9501\\
		UP& 4.95\%& 0.2892& 0.6394\\
		DDPG& $\mathbf{14.12}$\%& $\mathbf{0.5988}$ & $\mathbf{0.4913}$\\
		\bottomrule
	\end{tabular}
	
\end{table}
Overall, our cost DDPG strategy outperforms other strategies for the majority of the test trading period, though the advantage is not obvious at the beginning. The difference between cost DDPG strategy and others becomes obvious especially in the last few trading periods. Despite the excellent performance of the PAMR and OLMAR strategies from 2009 to 2010, their ultimate performance are worse than that of cost DDPG strategy due to the sharp decline after 2010. Both MDD and SR suggest the instability of these two strategies, which makes them vulnerable to the market full of risk. The performance metrics table(Table 1) indicates that the numerical results of cost DDPG strategy are the best among other strategies in profitability and stability aspects. In terms of CARR, the result of the DDPG strategy (14.12\%) is almost twice than that of the second largest benchmarks Anticor (7.94\%) over the test trading period. As for the risk measure, the cost DDPG strategy still has the best performance by holding the minimum MDD (0.4913), comparing with the UP strategy (0.6394) which is the second lowest. In terms of SR, cost DDPG strategy (0.5988) is nearly 33\% larger than that of Anticor strategy (0.3679), which is the second largest.

In conclusion, the aforementioned results demonstrate the outstanding and stable proﬁtability of DDPG strategy in comparison with other traditional strategies. 

\section{Conclusions and Future Work}
\subsection{Conclusions}
In this paper, we propose a novel pattern of DDPG reinforcement learning framework, including the LSTM critic network and the Fully Connected actor network. This DDPG strategy makes profits in our selected portfolio set --- 4 low volatility U.S stocks and 4 high volatility U.S stocks. Moreover, we use Compound Annual Return Rate, Sharpe Ratio, and Maximum Drawdown as performance metrics. In our experiment, the Cost DDPG strategy outperforms seven other traditional strategies in all three performance metrics. We also compare the performance of our DDPG strategy with and without transaction cost, and provide a plausible explanation to the result that DDPG strategy with transaction cost is more profitable and more robust than the Cost-free strategy.

Our strategy still has a number of drawbacks. First, we assume that there is no transaction slippage (Assumption 1), which is unrealistic in real market. Random transaction slippage happening might result in unexpected results in strategy reward. Second, we assume that each stock is available on any trading day (Assumption 6). Nevertheless, the stock might not be available sometimes, which will influence the return as well.

\subsection{Future Work}
For future work, we shall look into an improved DDPG strategy with transaction slippage in order to simulate the real market. The slippage might cause the strategy not to trade at the expected price, and then lead to the decline of profit. To solve this, we will try to modify the reward function so that the loss caused by slippage will be considered. Moreover, we will consider the situation that some stocks are not avaible in particular days. One possible solution to this problem is to expand our portfolio divisions, which will reduce the impact of non-tradable stocks on final return. Futhermore, our improved strategy will try to include short transactions by modifying the action space.

\end{document}